# Full *ab initio* atomistic approach for morphology prediction of hetero-integrated crystals: A confrontation with experiments

Sreejith Pallikkara Chandrasekharan[1], Sofia Apergi[1], Chen Wei[2], Federico Panciera[2], Laurent Travers[2], Gilles Patriarche[2], Jean-Christophe Harmand[2], Laurent Pedesseau[1,*], and Charles Cornet[1,*]

[1]Univ Rennes, INSA Rennes, CNRS, Institut FOTON - UMR 6082, F-35000 Rennes, France

[2]Centre de Nanosciences et de Nanotechnologies, CNRS, Université Paris-Saclay, 91120 Palaiseau, France

*Corresponding author: charles.cornet@insa-rennes.fr, laurent.pedesseau@insa-rennes.fr

## Abstract

Here, we propose a comprehensive first-principle atomistic approach to predict the Wulff-Kaischew equilibrium shape of crystals heterogeneously integrated on a dissimilar material. This method uses both reconstructed surface and interface absolute energies, as determined by density functional theory, to infer the morphology and wetting properties of Volmer-Weber islands over the whole range of accessible chemical potentials. The predicted equilibrium shapes of GaP crystals heterogeneously grown on Si, are found to be in good agreements with experimental observations performed by Transmission Electron Microscopy. Such method provides a tool for optimization of hetero-structured, multifunctional and smart materials and devices.

The central importance of interfaces between different materials and its key impact on devices performances was highlighted by *H. Kroemer* in his Nobel lecture [1]. Indeed, for all types of semiconductors, e.g. halide perovskites[2,3], II-VI [4], oxides[5], Ge[6], or III-Vs [7], surfaces and interfaces they form have very different physical properties than the ones of their corresponding bulk phases. While for years, surfaces and interfaces electronic states were considered as detrimental for semiconductor devices, recent works suggested that they could also give new opportunities for designing surface/interface-based quantum or classical devices[2,8,9].

In this context, the co-integration of dissimilar materials, also known as hetero-integration, has long been desired to develop high-quality, low-cost or multi-functional devices, especially for photonic, electronic, quantum and energy applications[8-14]. The quest of reduced dimensionality physical systems (2D, 1D or 0D) further increased the complexity of heterogeneous integration, such as in core-shell nanowires[15]. Amongst the various hetero-integration approaches, the III-V/Si monolithic integration is being considered for integrated photonics purposes or solar energy harvesting devices [10-14]. Recent experimental and theoretical research developments demonstrated that the hetero-epitaxial growth processes and subsequent physical properties of III-V epitaxial layers on Si are strongly related to the formation of 3D Volmer-Weber islands at the beginning of the crystal formation, which morphology directly impacts on the defects generation[14]. Especially, the formation of the hetero-interface was found to contribute significantly to the energy balance of the system [16,17].

The Wulff construction[18] is a method for predicting the shape of crystals from the surface energies. The construction especially allows to predict the Equilibrium Crystal Shape (ECS), as a consequence of the geometric minimization of the system's total energy, considering the surface energies contribution of the different crystal facets. The pertinence of this approach was confirmed experimentally in many different works[19]. Indeed, the fundamental understanding of the ECS provides substantial knowledge on numerous physicochemical properties, and is widely studied for determining crystal growth mechanisms[20-23], various catalytic [24] or luminescence properties [25].

Half a century later, the Wulff construction was generalized by considering the interaction between the studied crystal and the substrate on which it relies, thus forming an interface. This is a major extension to this approach. The equilibrium shape of a small particle on a solid substrate has been clarified by *R. Kaischew*[26], and further validated experimentally by *W. Winterbottom*[27]. Since then, several experimental studies have been conducted to observe the equilibrium crystal shapes from the various scanning microscopes, especially to improve the understanding of observed islands morphologies during the growth. These include growth of metal nanoparticles on metal oxides[28], semiconductor nanostructures[29,30], quantum dots[31,32], and perovskites[33]. In theory, based on a realistic description at the atomistic scale, a first-principles simulation can predict ECS. This was preliminary done within a Wulff-only approach. More recently, few studies have been carried out to implement at an atomic scale, a Wulff-Kaischew approach, e.g. for quantum dots[34], metal nanoparticles over metal oxides[35-38] and graphite's[39], metal oxides over perovskites[40], and semiconductor nanostructures[41]. Nevertheless, most studies used interface energy parameters derived from the relative energy approximations or the inverse Wulff-Kaischew approaches [42]. Among them, *Li et al.*[40] and *Dietze et al.*[36] considered the importance of determining absolute interface energies, and their changes with the chemical potential, although the description of a realistic interface remained complex. *L. G. Wang et al.* proposed an approach to account for interface energy contributions using the concept of wetting layer formation energy[43]. Indeed, the lack of accuracy in these interface energy values significantly affects the predicted morphological and wetting properties. This issue was methodologically solved in recent works, where absolute surface and interface energies were entirely determined for the specific case of GaP/Si with an unprecedented accuracy[14,16,17]; with this method, any surface (polar/nonpolar) or interface (abrupt, intermixed, or compensated) can now be modeled using a sufficiently large vacuum layer to minimize periodic boundary induced errors. Consequently, a comprehensive full *ab initio* atomistic description of the ECS, within the Wulff-Kaischew approach can now be explored.

In this article, we predict the equilibrium shape of crystals on solid surfaces within the Wulff-Kaischew approach, in a heterogeneous material system using a full *ab initio* atomistic description, based on absolute surfaces and interfaces energies computed by density functional theory (DFT). Equilibrium shapes are determined over the whole range of accessible chemical potentials. GaP on silicon monolithic integration is chosen as a specific case to assess the validity of the approach. A confrontation of the prediction with experimental observations performed by Transmission Electron Microscopy (TEM) is proposed for GaP/Si.

## Results

DFT calculations[44,45] were implemented in the SIESTA code[46-48] with basis set of finite-range numerical pseudo-atomic orbitals for the valence wave functions. Computational details are detailed in the methods.

Most of the absolute surface and interface energies used in this work are extracted from DFT calculations performed in ref. [16,17], and are reported in Supplementary Table S1. For one given surface or interface, the most stable atomic configuration was systematically selected for a given chemical potential, leading to a minimization of energies. For this specific study, we additionally calculated with the same methodology the surface energy of the {110} facets of GaP (Supplementary Fig. S1), being constant at around 33meV/Å², which is of the same order of magnitude as the one reported in the work of *Liu et al.*[49].

## Wulff equilibrium crystal shapes

The equilibrium Wulff shape, without considering the interaction with the substrate, is first considered. In a preliminary approach, the isometric Wulff construction enables the visualization of the crystal model by keeping the surface energies ($\gamma_s$) of all facets as constant, *i.e.*, $\gamma_s$ = 1 (arb. units). This isometric model allows to understand the arrangement of the considered stable facets that can form on the nanocrystal. The isometric Wulff shape for a GaP crystal is represented in Supplementary Fig. S2a & b. Supplementary Figure S2a shows the side view of the isometric Wulff shape, and the corresponding top view is shown in Supplementary Fig. S2b, with an orthogonal axis representation along the growth direction [001]. In this representation, the relative surface areas of the different facets are directly related to their index, and have no physical meaning. The considered crystal planes for determining the Wulff shape are obtained from experimental observations on epitaxial samples and include non-polar {001}, {110} and polar {111} A, {111} B, {2 5 11} A, {2 5 11} B, {114} A, {114} B facets [13,14,16,17,50]. Moreover, each A- and B-type facets (where A and B is defined in ref.[50]) have many sub-facets determined from the relation described in the Supplementary information. This methodology helps to predict the possible well-defined A & B polar facets, as shown in Supplementary Figure S2 (a) & (b). Figure S3 gives an illustration of the impact of the rotation matrix used to predict the configuration of the {2 5 11} facets. In this specific case, it leads to a total of four facets for each A and B in the top of the Wulff shape for the high index facets like {2 5 11}. From this isometric Wulff shape construction, the real equilibrium shape can now be considered taking into account surface energies determined by DFT (see Supplementary Table S1).

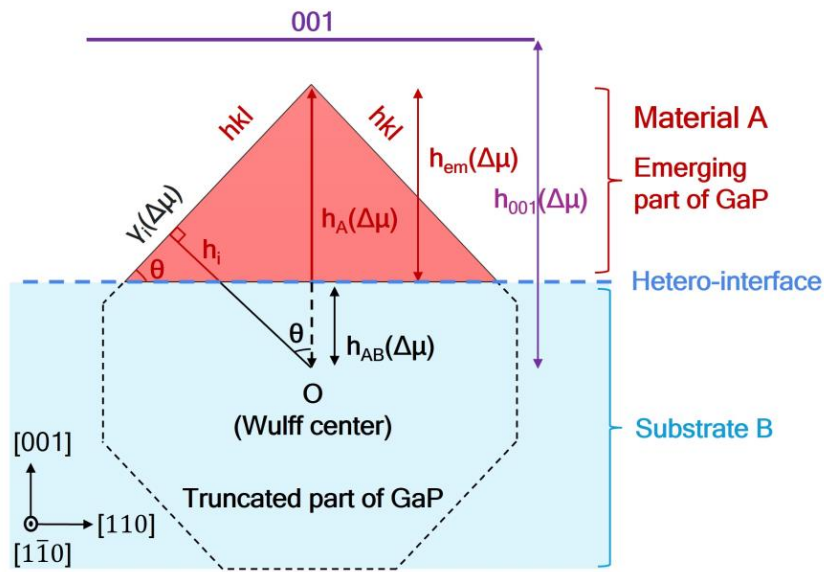

**Fig. 1. Methodology for analyzing Wulff-Kaischew approach.** Illustration of the methodology adapted to analyze the Wulff-Kaischew interaction of GaP nanocrystal on Si substrate.

Remarkably, equilibrium Wulff morphology depends on chemical potential variations within the nanocrystal since the surface energies, which themselves depend on the chemical potential, are pivotal in defining the equilibrium shape. This key observation was previously reported by *N. Moll et al.* in their study on the Wulff shape of GaAs[20]. Here, the variation of the phosphorus chemical potential is considered in the same way as in previous studies[16,17], over the thermodynamic stability range of GaP, between a Ga-rich limit (corresponding to the formation of a pure Ga phase) and a P-rich limit (corresponding to the formation of a pure P-phase, for GaP only) or a SiP-rich limit (corresponding to the formation of a pure SiP phase, for GaP/Si). The morphological changes of GaP crystals with respect to the variations of the phosphorus chemical potential are shown in Supplementary Fig. S5. It is demonstrated that, whatever the chemical potential, equilibrium shapes of GaP crystals are mostly governed by {110} and {111} facets, due to their low surface energies. With increasing phosphorus chemical potential, the {111} B facets develop and dominate whereas {110} facets tend to vanish, as shown in Supplementary Fig. S5. It is also noticed that the top surface facets are a complex mixture of polar {111}, {2 5 11}, {114} and non-polar {001}, {110} facets, whose energies are strongly dependent of the chemical potential variations. This phenomenon is a direct consequence of the full *ab initio* atomic scale description of the considered reconstructed surface stoichiometries and cannot be accurately predicted by non-atomic scale modelling.

## Wulff-Kaischew equilibrium crystal shapes

Interaction with the substrate is now considered in the Wulff-Kaischew approach. Figure 1 illustrates the Wulff-Kaischew construction for a GaP crystal (material A, filled in red), in equilibrium on a Si (001) surface (substrate B). In this drawing, the black dotted line represents the ECS of the GaP crystal without any interaction with the substrate, the Wulff center of the crystal is O and $h_i$ is the distance of any facet hkl from O, associated to the surface energy ($\gamma_i$), $h_A$ is the distance between the top of the emerging crystal and O. Anisotropy in the surface energy can cause $h_A$ fluctuations at heights below $h_{001}$, which corresponds to the equilibrium distance between the {001} facet and O along the growth direction [001], resulting in the disappearance of the {001} facets. $h_{AB}$ is the distance between O and the substrate surface, and theta ($\theta$) is the angle between the substrate B surface (dashed blue line) and the facet hkl of the material A. The height of the GaP island (red color) grown on top of the Si is the emerging height ($h_{em}$). In addition to all the previously used surface energies, the Wulff-Kaischew approach supposes the quantitative knowledge of the interface energy and of the substrate surface energy. For the GaP/Si interface, the most stable GaP/Si atomic configuration is chosen (i.e. the Ga-charge compensated interface, described in ref. [16]). For the surface energy, Si surfaces passivated by monoatomic layers of Ga or P are considered [17], depending on the chemical potential variations.

The Wulff-Kaischew ECS can thus be constructed, with the following equation : [52],

$$\lambda = \frac{2\gamma_A - \beta}{h_{em}} = \frac{\gamma_i}{h_i}, \tag{1}$$

where, $\lambda$ is a size constant, $\gamma_A$ is the surface energy of the basal face of material A before adhesion into substrate, $\beta$ is the adhesion energy of the system, $h_{em}$ is the emerging height from the substrate surface, $\gamma_i$ and $h_i$ are the surface energies and heights of all singular facets i of the free non-supported GaP crystal, respectively, as shown in Fig. 1. Importantly, all these terms depend on the variations of chemical potentials (Δμ). With this Wulff-Kaischew equation, wetting properties of the system are discussed in the Supplementary information (see details in Supplementary Fig. S6, S7, and Supplementary note).

Following the Wulff-Kaischew equation (1), the ECS of GaP crystals on Si substrate can be calculated, based on absolute GaP and Si surface and GaP/Si interface energies determined from DFT and described above for the whole range of accessible phosphorus chemical potential. Figure 2 represents the full *ab initio* ECS of GaP islands on a Si substrate for various chemical potentials between the Ga-rich and SiP-rich thermodynamic boundaries. The shape of these islands varies with the different thermodynamic growth conditions, from pyramidal (extreme values of the chemical potential) to pyramidal truncated (intermediate values of the chemical potential). In Ga-rich conditions (top left), {111} A facets contribute more to the ECS, while in SiP-rich conditions (bottom left) {111} B facets become prominent. In between, with the variation of chemical potentials, different other stable facets, including {001}, {110} {2 5 11}, and {114}, are showing up on the GaP surface, with corresponding color codes for each, as illustrated in Fig. 2. From the above analysis, a relative description of the contributions of surface area for each facet family in all the intervals can be generated, as shown in Supplementary Fig. S8. More specifically, between $\Delta\mu_P$ = -0.90 and -0.70 eV, {001} and {2 5 11} B facets are visible, because respective surface energies of both facets lie in the [55-60 meV/Å²] range. The arising of {114} A facets are visible at $\Delta\mu_P$ = -0.70 and -0.60 eV. In this interval, {2 5 11} B facets are also visible, but progressively disappear with the emergence of the {2 5 11} A facets. The {001} facet also progressively disappear. After this, between $\Delta\mu_P$ = -0.50 eV and -0.30 eV, the dominant contributions are {2 5 11} A and {111} facets. This evolution ends with {111} facets mostly at -0.196 eV. Overall, the ECS Wulff-Kaischew morphology maintains a rectangular base shape on the whole range of chemical potential. DFT analysis shows the important role of {111} and {110} facets in establishing the shape. Nevertheless, differences in the surface energies result in the change of shape and size of Wulff-Kaischew morphologies, resulting in [1$\bar{1}$0]-elongated islands in Ga-rich conditions, and [110]-elongated islands in SiP-rich conditions. In general, it is interesting to note that when considering experimental observations III-V/Si islands, both rectangular-shaped along the [110] or [1$\bar{1}$0] directions [53] and square-shaped islands [14] are observed, depending on the materials considered, or the growth conditions. This is briefly discussed in Supplementary information (see Supplementary Fig. S9 and Supplementary note).

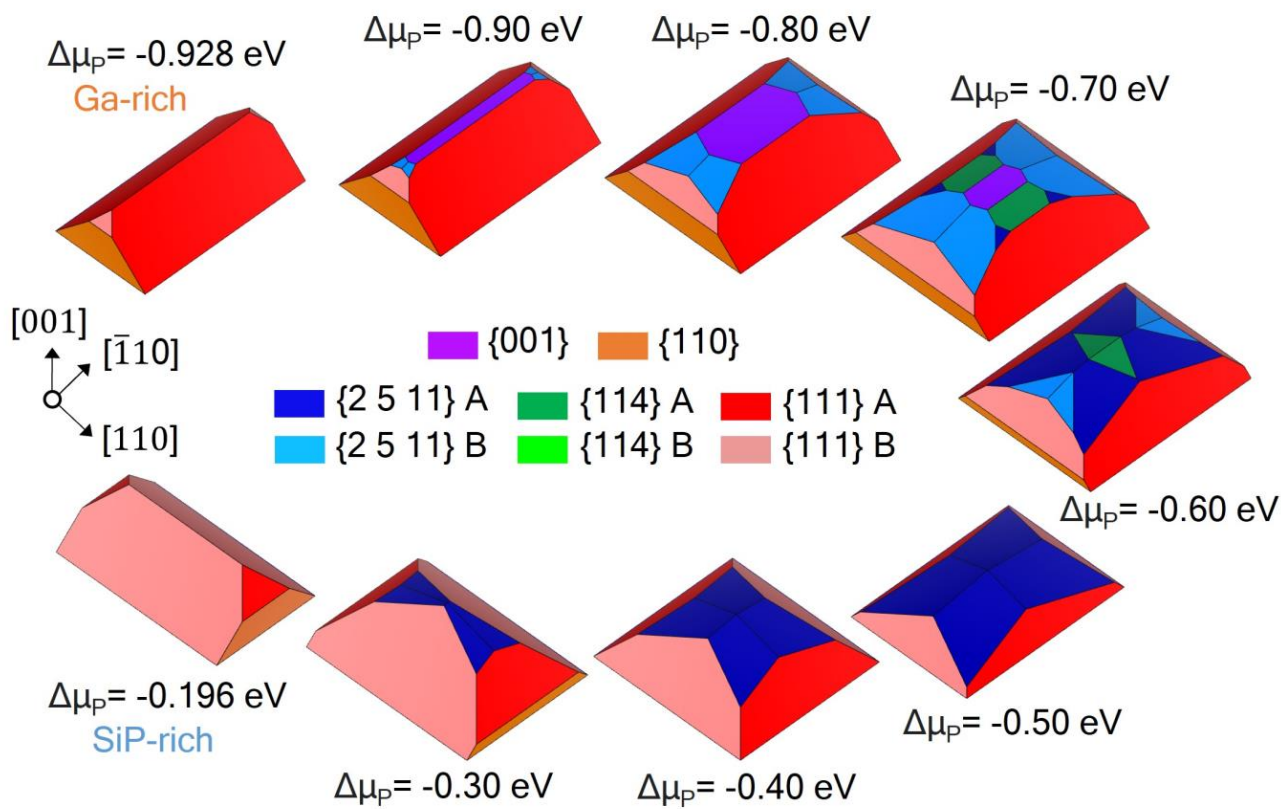

**Fig. 2. Equilibrium shape evolution of GaP on Si (001).** The Wulff-Kaischew shape evolution of GaP grown on Si (001) substrate based on variations of chemical potentials in the thermodynamic stability range. The respective color code for each facet is illustrated within the figure.

Experimental validation of such an ECS determination using a full *ab initio* method is not straightforward, as epitaxial growth of GaP on Si does not usually happen in thermodynamic equilibrium conditions, and of course not at 0K, and islands are subjected to various kinetically-limited processes. A statistical approach is thus followed. Here, few nm-thick GaP was grown on Si(001) and observed *in situ* by plan-view TEM in nucleation-driven growth conditions[54]. Growth processes and microscopy are discussed in methods. Fig. 3a shows many small islands on a 400x400 nm$^2$ surface, with an island density of 5.7x10$^{11}$ cm$^{-2}$ and a coverage of 39%, and a mean lateral dimension of 9 nm (Supplementary Fig. S10)

Full *ab initio* atomistic approach for morphology prediction of hetero-integrated crystals: a confrontation with experiments in good agreement with previous studies[14]. At this stage of growth, some of these islands already started to coalesce to form larger islands. So, for the purposes of this study, where individual islands before coalescence are under scrutiny, islands with an area > 250 nm² have been excluded. Within the surface, a visible elongation of individual islands is observed towards [1$\bar{1}$0] or [110] directions. This is attributed to the nucleation of GaP islands of different phases (referred as A and B) on Si terraces having different polarities, in good agreement with the nucleation-driven antiphase domain distribution scenario[54]. Note that TEM does not allow to discriminate the [110] and [1$\bar{1}$0] directions within III-V islands. Therefore, islands are assumed to be all elongated along either only the [110] direction or only along the [1$\bar{1}$0] direction.

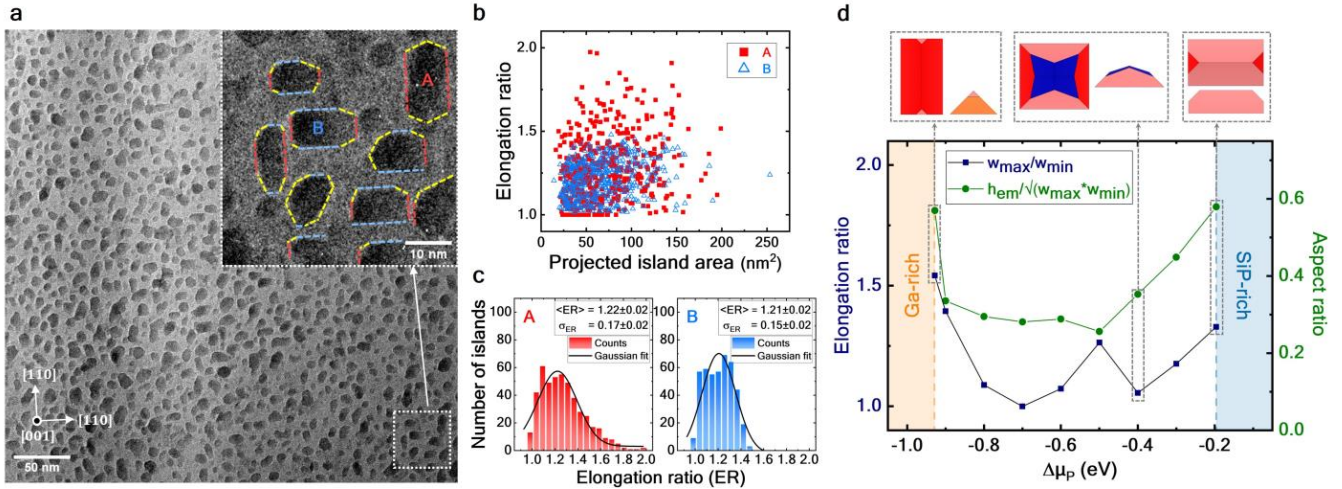

**Fig. 3. Morphological and structural analysis of GaP islands on Si (001). a** 400x400 nm² TEM plan-view image of GaP islands on Si (001) showing numerous 3D islands with a highlighted 50x50 nm² surface with identification of island orientation, and where A and B represent the two phases of the crystal. Crystallographic directions of the Si substrate are given. **b** Elongation ratios of counted islands from the 400x400 nm² image corresponding to the A and B phases along with the quantitatively measured area of projected islands. **c** Distribution of elongation ratios of individual islands belonging to the A and B phases with a gaussian fitting. **d** DFT predicted elongation and aspect ratios of GaP/Si islands between the Ga-rich and SiP-rich thermodynamic limits of stability along with the shape of islands in three different intervals.

An in-depth analysis is performed on the 50x50 nm$^2$ area highlighted in Fig. 3a inset. Edges of islands were identified manually, and their angles with the in-plane substrate crystallographic orientation measured (See Supplementary Fig. S11, Table S2, and Supplementary note). The study reveals prominent contributions of edges along [110] and [1$\bar{1}$0] directions (blue and red dotted lines in Fig. 3a, as well as [250] (or [130]) additional edges (yellow dotted lines in Fig. 3a). While these additional edges (yellow dotted lines) are not predicted by DFT at the thermodynamic equilibrium, it is worth mentioning that the in-plane directions observed can intersect the {111}, {110}, {114} or {2 5 11} facets identified as stable by the DFT, and having the largest contribution to the ECS.

Figure 3b represents the statistical distribution of measured in-plane elongation ratio for A- and B-phase islands (in red and blue respectively). Here, the elongation ratio is defined as $\frac{w_{max}}{w_{min}}$, where $\frac{w_{max}}{w_{min}}$ are the largest (smallest) lateral width of the islands. Intrinsically, the statistical analysis traduces the bidomain behavior in the elongation ratios of the two populations (A and B). The average values are identified with a Gaussian fit at around 1.22±0.02 and 1.21±0.02 for A and B islands respectively. The results are almost the same, as shown in Fig. 3c, although the A islands have a slightly larger inhomogeneous broadening, which could be induced by kinetically limited surface processes due to the well-known difference between the structure of Si monoatomic steps along the [110] or [1$\bar{1}$0] directions and the associated anisotropic adatom diffusion processes.

Figure 3d summarizes the prediction of the full *ab initio* ECS prediction of Fig. 2 over the whole thermodynamic stability range. Elongation and aspect ratios of GaP islands on Si substrates are plotted versus the chemical potential variations. In this plot, the elongation ratio is defined again as $\frac{w_{max}}{w_{min}}$ (see details in methods). [110] and [1$\bar{1}$0] are not differentiated, for a better comparison to experiments. Remarkably, DFT results predict an elongation ratio in the range of 1.0 to 1.5 over the whole range of chemical potentials, precisely matching the Gaussian distribution of the experimental measurements. Going beyond in the analysis is not reasonable because it is not possible to establish a direct straightforward equivalency between experimental atom fluxes and the defined chemical potentials.

Besides, the theoretical value of the aspect ratio, defined as $\frac{h_{em}}{\sqrt{(w_{max}*w_{min})}}$ (detailed in methods), lies typically between 0.3 and 0.6. We were unable to compare this range of values with the experimental aspect ratios corresponding to Fig. 3a, as the sample was not observed in cross-section. However, these values are consistent with the 0.3 to 0.8 range observed in previous cross-sectional TEM or plan-view STM (Scanning Tunneling Microscopy) studies on III-V/Si individual islands[14,42,55,56]. Furthermore, we analyzed two distinct regimes: (i) the end of GaP growth and (ii) annealing under P for 1 minute. Our observations suggest that the system does not evolve significantly upon annealing, and is therefore 'close to equilibrium' here (See Supplementary Fig. S12 and Supplementary note. Overall, even if a direct comparison between theoretical predictions and experiments remains delicate, due to non-equilibrium conditions, inability to determine anisotropic kinetically-limited adatoms diffusion processes related to the nature of the Si steps or to the residual miscut, the morphology of islands predicted from the full *ab initio* ECS is in remarkable agreement with the experimental statistical analysis presented here, confirming the pertinence of the full *ab initio* Wulff-Kaischew approach.

## Discussion

In this study, we have used absolute surface and interface energy values determined from density functional theory to achieve a full *ab initio* description of the Wulff-Kaischew equilibrium shapes of crystals over a substate. This atomistic approach applied on heterogeneous materials systems, such as III-V/Si, provides comprehensive understandings on the possible nanocrystal facets formation during the hetero-integration of the crystal on a dissimilar substrate. It also predicts the morphological changes of the crystal due to the variations of thermodynamic conditions (e.g. chemical potentials). The morphologies predicted by the ECS atomistic approach are found to be in excellent agreement with *in situ* TEM observations. Such a full *ab initio* ECS Wulff-Kaischew approach is expected to be of great interest for the understanding of growth processes and physical properties of hetero-integrated materials and devices.

## Methods

### Computational details

We performed all the DFT calculations as implemented in the SIESTA code[46-48] with basis set of finite-range numerical pseudo-atomic orbitals for the valence wave functions. Generalized Gradient Approximation functional in the PBE form[57] of the Troullier-Martins pseudopotentials[58] was utilized as an exchange-correlation functional. The integration of the Brillouin zone was accomplished via 4×4×1 Monkhrost-Pack k-points[59]. A vacuum thickness of 150 Å spacing is introduced in the vertical direction of the substrate or in the z-direction to eliminate the interactions between periodic images.

### Analyzing the Wulff-Kaischew interaction

In this section, we discuss the method based on full *ab initio* to obtain the Wulff-Kaischew shape of heterogeneous materials. To this aim, we consider a hetero-integration scenario of GaP crystal (A) on Si (001) substrate (B), where, Si is passivated with group III or V elements to validate the wetting conditions[17]. In Figure S6(a) and (b), the Wulff center of the crystal is expressed as O and $h_i$ is the height of any facet hkl from O with the surface energy ($γ_i$), $h_A$ is the distance between the top of the emerging crystal and O. The $h_{AB}$ is the height from O to the substrate surface, and θ is the angle between the substrate surface to the facet hkl. As a result of the anisotropy in surface energies, this $h_A$ can be lower than height to the growth direction [001] ($h_{001}$), as shown in Supplementary Fig. S6a. Also, when a new i height $h_i$' appears over the top facet, $h_a$ changes to $h_a$' with a new angle θ'. The height of the GaP grown on top of the Si is known as emerging height ($h_{em}$), emerged area indicated in red color, as shown in Supplementary Fig. S6a. This emerged height can be determined from the relation, $h_{em} = h_A - h_{AB}$, while $h_{AB}$ can be calculated from[60],

$$\frac{h_{AB}}{h_A} = \left(1 - \frac{β_i}{γ_A}\right), \quad (2)$$

simply,

Full *ab initio* atomistic approach for morphology prediction of hetero-integrated crystals: a confrontation with experiments

$$\frac{h_{AB}}{h_A} = \frac{\gamma_{AB} - \gamma_B}{\gamma_A}, \tag{3}$$

where, $\frac{h_{AB}}{h_A}$ denotes the truncation ratio, discussed in the next section. $\gamma_A$ is the surface energy of the basal face of material A before adhesion into substrate, $\gamma_B$ is the surface energy of the substrate, $\gamma_{AB}$ is the interface energy, and $\beta_i$ is the adhesion energy of the system (depending on i). Moreover, the change in $h_{em}$ leads to a morphology change of the top crystal surface of GaP. Although, top surface changes with the orientation of low-surface energy facet, the interface parameter $h_{AB}$ changes also with change in interface energy of the system.

To understand the dimensional details of the Wulff-Kaischew shape, fundamental understanding of its elongation and aspect ratios are required. These parameters define the change in shape of the Wulff-Kaischew approach. In this study, significant contributions to the area of the shapes comes from {111} and {110} GaP crystal facets, so we considered the width along both elongation directions ([1$\bar{1}$0] and [110]) to obtain the elongation ratio. This minimum and maximum widths along these directions have been calculated from the methodology as shown in Supplementary Fig. S6b, where we introduced three new parameters $\frac{w_0}{2}$, $\frac{w}{2}$, and x. Here, $\frac{w_0}{2}$ is the half width of the Wulff shape of a crystal along the [110] or [1$\bar{1}$0] direction from the Wulff center, without any interaction with the substrate. $\frac{w}{2}$ is the half distance of the Wulff-Kaischew shape at the interface, along the same direction. and x is difference $\frac{w_0}{2}$ - $\frac{w}{2}$.

To calculate these new width parameters, we define the following relationships,

$$\frac{w_0}{2} = h_i / \sin\theta \tag{4}$$

$$x = h_{AB} / \tan\theta \tag{5}$$

$$\frac{w}{2} = \frac{w_0}{2} - x \tag{6}$$

then the actual width of the new truncated shape equals to $2 * (\frac{w}{2})$ = w. Based on the above, the elongation ratio can be defined as,
13

$$\text{elongation ratio} = \frac{w_{max}}{w_{min}} \tag{7}$$

Similarly, the aspect ratio, i.e. the relationship between height and width, is defined as,

$$\text{aspect ratio} = \frac{h_{em}}{\sqrt{(w_{max}*w_{min})}} \tag{8}$$

# Experimental details

The GaP sample discussed in this article were grown inside a Titan environmental transmission electron microscope (TEM) prototype designed in collaboration with FEI and installed at Ecole Polytechnique in Palaiseau (France). It operates at 300 keV and the image is corrected from geometrical aberrations. This equipment allows us to perform molecular beam epitaxy (MBE) inside the TEM. Ga and $P_2$ molecular fluxes were generated from collimated effusion sources to minimize the deposition in the column of the microscope. *In situ* and real-time observation of growth by TEM was possible by using a MEMS (micro-electromechanical system) chip mounted on the extremity of the custom sample holder, as substrates. The chips are fabricated at C2N from a silicon-on-insulator substrate consisting of a highly doped Si (001) device layer, a $SiO_2$ buried oxide layer and a Si handling wafer. Thin membranes with a surface area of 100 x 100 µm² are defined by photolithography, plasma dry etching and chemical wet etching to locally remove the handling wafer and oxide layer from the backside. Its final thickness is about 60 nm, and it is, therefore, transparent to the TEM electron beam. On the front side, four arms are defined by lithography and dry etching to hold the membrane suspended above the via-hole. These arms are designed to minimize the distortion of the membrane during heating and to maintain a homogeneous temperature on it. The heating is obtained by Joule effect - a voltage drop is applied between two pairs of arms to flow current through the membrane- and temperatures of up to 900°C are easily achieved. The actual temperature is pre-calibrated by measuring the shift of Si Raman peak as a function of the applied voltage. The Si membrane is deoxidized with 5% dilute HF solution just before its loading into the TEM column and a further oxide thermal desorption is effectuated *in situ* at 900°C.

**Data availability**

All data needed to evaluate the conclusions in the paper are present in the paper and/or the Supplementary Information. The other data that supports the findings of this study are available from the corresponding author upon request.

**Acknowledgments**

This research was supported by the French National Research NUAGES Project (Grant no. ANR-21-CE24-0006). DFT calculations were performed at FOTON Institute OHM - INSA Rennes, and the work was granted access to the HPC resources of TGCC/CINES under the allocation A0120911434, A0140911434, A0160911434 made by GENCI.

## Author contributions

S.P.C. conducted the investigation, carried out formal analysis, curated the data, and prepared the original draft. S.A. contributed to data analysis and manuscript review and editing. C.W., F.P., and J.-C.H. prepared the samples. G.P., J.-C.H., L.T., and S.P.C. carried out the material growth and performed *in situ* TEM observations. S.P.C. conducted data analysis. C.C. and L.P. conceived the study, performed methodology development, supervised the project, validated the results, contributed to data analysis, manuscript review and editing, and contributed to project administration and conceptualization. C.C. acquired funding, and L.P. was responsible for resource management.

## Competing interest

The authors declare no competing interests.

## Additional information

**Supplementary information** The online version contains supplementary material available at

**Correspondence** and requests for materials should be addressed to Charles Cornetand Laurent Pedesseau.